\documentclass[conference,10pt,twoside]{IEEEtran}
\normalsize
\newif\ifonecolumn
\IEEEoverridecommandlockouts
\usepackage[final]{graphicx}
\usepackage{times}
\usepackage{amsfonts}
\usepackage{epsfig}
\usepackage{amssymb}
\usepackage{amsmath}
\usepackage{citesort}
\usepackage{psfrag}
\usepackage{balance}
\usepackage{comment}
\usepackage{bm}
\usepackage{color, xcolor,colortbl}

\newcommand{\beq}{\begin{equation}}
\newcommand{\eeq}{\end{equation}}
\newcommand{\ben}{\begin{enumerate}}
\newcommand{\een}{\end{enumerate}}
\newcommand{\bit}{\begin{itemize}}
\newcommand{\eit}{\end{itemize}}

\newtheorem{example}{Example}
\newtheorem{lemma}{Lemma}

\newtheorem{theorem}{Theorem}

\newtheorem{proposition}{Proposition}

\DeclareMathOperator*{\argmax}{arg\,max}
\DeclareMathOperator{\spn}{range}

\definecolor{LightCyan}{rgb}{0.88,1,1}

\makeatletter
\newenvironment{mysmallmatrix}{%
    \let\saved@CT@row@color\CT@row@color
    \begin{smallmatrix}%
}{%
    \end{smallmatrix}%
    \global\let\CT@row@color\saved@CT@row@color
}
\makeatother

\begin{document}
\title{On the Minimum Distance of Array-Based Spatially-Coupled Low-Density Parity-Check Codes}

\author{\IEEEauthorblockN{Eirik Rosnes}\\
\vspace{-4mm} \IEEEauthorblockA{Department of Informatics,
University of Bergen, N-5020 Bergen, Norway, and the Simula Research Lab.\\ Email:
eirik@ii.uib.no}
\thanks{This work was partially funded by the Norwegian-Estonian Research Cooperation Programme (grant EMP133), the Research Council of Norway (grant 240985/F20), and by Simula@UiB.}}

\maketitle

\maketitle

\begin{abstract}
An array low-density parity-check (LDPC) code is a quasi-cyclic LDPC code specified by two integers $q$ and $m$, where $q$ is an odd prime  and $m \leq q$. The exact  minimum distance, for small $q$ and $m$, has been calculated, and tight upper bounds on it for $m \leq 7$ have been derived.  %
In this work, we study the minimum distance of the spatially-coupled version of these codes. 
In particular, several tight upper bounds on the \emph{optimal} minimum distance for coupling length at least two and $m=3,4,5$, that are independent of $q$ and that are valid  for all values of $q \geq q_0$ where $q_0$ depends on $m$, are presented. Furthermore, we show by exhaustive search that by carefully selecting the \emph{edge spreading} or \emph{unwrapping}  procedure, the minimum distance (when $q$ is not very large) can be significantly increased, especially for $m=5$. 
\end{abstract}

\section{Introduction}

In this paper, we consider array-based spatially-coupled low-density parity-check (SC-LDPC) codes as introduced in \cite{bal14} and subsequently studied in \cite{mit14} from the perspective of \emph{absorbing sets}.  Array LDPC codes were originally proposed by Fan in \cite{fan00} and are specified by two integers $q$ and $m$, where $q$ is an odd prime and $m \leq q$. 

Spatial coupling was first introduced in the coding theory literature by  Felstr\"{o}m and Zigangirov in \cite{fel99}, where they proposed 
convolutional LDPC or SC-LDPC codes. These codes have very good belief propagation (BP) decoding thresholds due to the phenomenon of \emph{threshold saturation}. For instance, in \cite{kud13}, it was shown that for binary memoryless channels, the BP decoding threshold \emph{saturates} to the maximum \emph{a posteriori} decoding threshold of the underlying ensemble. %

Since the original work by Fan \cite{fan00}, several authors have considered the \emph{structural} properties (including the minimum distance $d_{\rm min}$) %
of array LDPC codes (see, e.g., \cite{mit02,yan03,sug08,esm09,ros14,dol10}). For high rate and
moderate length, these codes perform well under iterative decoding, and they are also well-suited for practical implementation due to their regular structure \cite{olc03,bha05}.

In this work, we present several tight upper bounds on the \emph{optimal} minimum distance of array-based SC-LDPC codes  for coupling length at least two and $m=3,4,5$, that are independent of $q$ and that are valid  for all values of $q \geq q_0$ where $q_0$ depends on $m$. Also, these bounds (and intermediate results in their proofs) 
can in some cases be used to quickly remove large parts of the search space when searching for \emph{optimal} \emph{cutting vectors}. For small values of $q$ ($m=3,4,5$) we present the results of an exhaustive search over all cutting vectors, 
showing that a careful selection can increase the minimum distance significantly, especially for $m = 5$.

We remark that  the iterative decoding performance of these codes, for instance, on additive white Gaussian noise channels, is typically dominated by minimum absorbing sets and their multiplicities. However, improved decoding methods and ultimately maximum-likelihood decoding will not be trapped in absorbing sets and thus overcome the shortcomings of standard iterative decoding, in which case the minimum distance (and its multiplicity) will become  an important performance metric.

\section{Array LDPC Codes} \label{sec:arrayLDPC}

The array LDPC code $\mathcal{C}(q,m)$, with parameters $q$ and $m$, has length $q^2$ and can be defined by the parity-check matrix
\begin{equation} \label{eq:pcmatrix}
\bm{H}(q,m) = \left[ \begin{matrix}
\bm{I} & \bm{I} & \bm{I} & \cdots & \bm{I} \\
\bm{I} & \bm{P} & \bm{P}^2 & \cdots & \bm{P}^{q-1} \\
\bm{I} & \bm{P}^2 & \bm{P}^4 & \cdots & \bm{P}^{2(q-1)} \\
& & \vdots  & & \vdots \\
\bm{I} & \bm{P}^{m-1} & \bm{P}^{2(m-1)} & \cdots & \bm{P}^{(m-1)(q-1)} \end{matrix} \right]
\end{equation}
where $\bm{I}$ is the $q \times q$ identity matrix and $\bm{P}$ the $q \times q$ permutation matrix
\begin{displaymath}
\bm{P} = \left[ \begin{matrix}
0 & 0 & \cdots & 0 & 1 \\
1 & 0 & \cdots & 0 & 0 \\
0 & 1 & \cdots & 0 & 0 \\
& \vdots & &  \vdots & \\
0 & 0 & \cdots & 1 & 0 \end{matrix} \right].
\end{displaymath}
Since the number of ones in each row of the matrix in (\ref{eq:pcmatrix}) is $q$ and the number of ones in each column is $m$, the array LDPC codes are $(m,q)$-regular codes. Furthermore, its %
rank is $qm-m+1$, from which it follows that the dimension of $\mathcal{C}(q,m)$ is $q^2-qm+m-1$. 

In \cite{yan03}, a new representation for $\bm{H}(q,m)$ was introduced. In particular, since each column of the parity-check matrix $\bm{H}(q,m)$ has $m$ blocks and each block is a permutation of $(1,0,0,\dots,0,0)^T$, where $(\cdot)^T$ denotes the transpose of its argument, we can represent each column as a vector of integers between $0$ and $q-1$, where
\begin{equation} \label{eq:rep}
i \triangleq \left(\overbrace{0,\dots,0}^{i},1,\overbrace{0,\dots,0}^{q-i-1} \right)^T
\end{equation}
i.e., the $1$-positions are associated with the integers modulo $q$. %
 Furthermore, it follows from (\ref{eq:pcmatrix}) and the integer representation in (\ref{eq:rep})  that any column in an array LDPC code parity-check matrix is of the form 
\begin{equation} \label{eq:form}
(x,x+y,x+2y,\dots,x+(m-1)y)^T \bmod q
\end{equation}
where $(x,y) \in [q]^2$ and, for notational convenience, for any positive integer $L$, $[L] \triangleq \{0,1,\dots,L-1\}$. Thus, a column can be specified by two integers $x$ and $y$. Also, note that since there are $q^2$ distinct columns in an array LDPC code parity-check matrix, any pair $(x,y) \in [q]^2$ %
specifies a valid column.

In the following, it is also convenient to consider the parity-check matrix in (\ref{eq:pcmatrix}) as an $m \times q$ array of $q \times q$ permutation matrices with \emph{row group} indices $i \in [m]$ and \emph{column group} indices $j \in [q]$, from which it follows that each column can be uniquely indexed by the pair $(j,k)$, where $j \in [q]$ is the index of the column group and $k \in [q]$ is the index within a column group. Likewise, each row can be uniquely indexed by the pair $(i,k)$, where $i \in [m]$ is the index of the row group and $k \in [q]$ is the index within a row group.

\section{Array-Based  SC-LDPC Codes} \label{sec:SCarrayLDPC}

Array-based SC-LDPC codes can be constructed from array LDPC codes by a special type of \emph{edge spreading} or \emph{unwrapping} procedure \cite{fel99,pus11} specified by a ``cutting'' vector $\boldsymbol{\zeta} = (\zeta_0,\dots,\zeta_{m-1})$, where $0 \leq \zeta_0 < \zeta_1 < \cdots < \zeta_{m-1} \leq q$. The purpose of the cutting vector is to generate two parity-check matrices $\bm{H}_0$ and $\bm{H}_1$, each of size $m q \times q^2$ and initially filled with zeros, as follows:
\begin{itemize}
\item For each $\zeta_i$, $i \in [m]$, the $q \times q$ permutation matrices in row group $i$ and column group $j$, $j < \zeta_i$, of $\bm{H}(q,m)$ are copied into the corresponding positions in $\bm{H}_0$.
\item Similarly, for each $\zeta_i$, $i \in [m]$, the $q \times q$ permutation matrices in row group $i$ and column group $j$, $\zeta_i \leq j < q$, of $\bm{H}(q,m)$ are copied into the corresponding positions in $\bm{H}_1$.
\end{itemize}

\begin{example}
If $q=5$, $m=3$, and $\boldsymbol{\zeta}=(1,2,4)$, the matrices $\bm{H}_0$ and $\bm{H}_1$ become
\begin{displaymath}
\bm{H}_0 = \left[ \begin{matrix}
\bm{I} & \bm{0} & \bm{0} & \bm{0} & \bm{0} \\
\bm{I} & \bm{P} & \bm{0} & \bm{0} & \bm{0} \\
\bm{I} & \bm{P}^2 & \bm{P}^4 & \bm{P} & \bm{0}
\end{matrix} \right]
\end{displaymath}
and
\begin{displaymath}
\bm{H}_1 = \left[ \begin{matrix}
\bm{0} & \bm{I} & \bm{I} & \bm{I} & \bm{I} \\
\bm{0} & \bm{0} & \bm{P}^2 & \bm{P}^3 & \bm{P}^4 \\
\bm{0} & \bm{0} & \bm{0} & \bm{0} & \bm{P}^3
\end{matrix} \right].
\end{displaymath}
\end{example}

For a given positive integer $L$ (the coupling length) and a cutting vector $\boldsymbol{\zeta}$, an array-based SC-LDPC code is defined by the parity-check matrix
\begin{equation} \label{eq:SCpcmatrix}
\bm{H}(q,m,L,\boldsymbol{\zeta}) = \left[ \begin{matrix}
\bm{H}_0 & & & \\
\bm{H}_1 & \bm{H}_0 & & \\
 & \bm{H}_1 & \smash{\ddots} & \\
& & \smash{\ddots} & \bm{H}_0 \\
& & & \bm{H}_1 \end{matrix} \right]
\end{equation}
of size $(L+1) m q \times L q^2$.

As for the uncoupled case, it is convenient to consider the convolutional  parity-check matrix in (\ref{eq:SCpcmatrix}) as a $(L+1)m \times Lq$ array of $q \times q$ permutation and all-zero matrices in which $m q$ consecutive rows (resp.\ $q^2$ consecutive columns) are referred to as a row (resp.\ column) \emph{section}. Within each row (resp. column) section, the rows (resp.\ columns) correspond to a row (resp.\ column) group. In summary, each column can be indexed by a triple $(l,j,k)$, where $l \in [L]$ is the column section index, $j \in [q]$ is the column group index within column section $l$, and $k \in [q]$ is the column index within column group $j$ in column section $l$. Similarly, each row can be indexed by a triple $(l,i,k)$, where $l \in [L+1]$ is the row section index, $i \in [m]$ is the row group index within row section $l$, and $k \in [q]$ is the row index within row group $i$ in row section $l$.

Note that as for the uncoupled matrix in (\ref{eq:pcmatrix}), all columns in $\bm{H}(q,m,L,\boldsymbol{\zeta})$ contain exactly $m$ ones and each row in row sections $1,2,\dots,L-1$ contains exactly $q$ ones. However, the rows in row sections $0$ and $L$ can have fewer than $q$ ones.

In the following, we will denote the binary linear code defined by the coupled parity-check matrix $\bm{H}(q,m,L,\boldsymbol{\zeta})$ in (\ref{eq:SCpcmatrix}) by $\mathcal{C}(q,m,L,\boldsymbol{\zeta})$ and its corresponding minimum (resp.\ stopping) distance by $d(q,m,L,\boldsymbol{\zeta})$ (resp.\ $h(q,m,L,\boldsymbol{\zeta})$).

It was shown in \cite[Theorem 5]{mit14} that the minimum distance  $d(q,m,L,\boldsymbol{\zeta})$ of $\mathcal{C}(q,m,L,\boldsymbol{\zeta})$ is bounded below by the minimum distance $d(q,m)$ of $\mathcal{C}(q,m)$ for all cutting vectors $\boldsymbol{\zeta}$.

Now, let us define the optimal minimum distance of an array-based SC-LDPC code as the highest minimum distance over all possible cutting vectors as the coupling length $L$ tends to infinity, i.e., 
\begin{displaymath}
d_{\rm opt}(q,m)  \triangleq \lim_{L \to \infty} d_{\rm opt}(q,m,L)
\end{displaymath}
where 
\begin{displaymath}
d_{\rm opt}(q,m,L) \triangleq \max_{\boldsymbol{\zeta}} d(q,m,L,\boldsymbol{\zeta}).
\end{displaymath}

\section{Upper Bounds on  $d_{\rm opt}(q,m,L)$}

In this section, we derive upper bounds on $d_{\rm opt}(q,m,L)$ for $L \geq 2$ and $m=3,4,5$, that are independent of $q$ and that hold for all values of $q \geq q_0$ where $q_0$ depends on $m$.

Now, define the \emph{range} of a sorted (in nondecreasing order) sequence of column group indices $\bm{j}^{\rm gr} = (j_0^{\rm gr},\dots,j_{n-1}^{\rm gr})$ of length $n$ as 
\begin{displaymath}
\min_{l \in [n]} \left( j_{l}^{\rm gr}-\left(j_{(l+1) \bmod n}^{\rm gr} -  q \left \lceil \frac{n-1-l}{n-1} \right \rceil \right) \right) +1.
\end{displaymath}
 Furthermore, define
\begin{align}
\epsilon(\boldsymbol{\zeta}) &\triangleq  \max_{j \in [m-1]} \left( \left(\zeta_{(j+1) \bmod m}-\zeta_{j}\right)  \bmod q \right) \notag \\
j_{\rm max}(\boldsymbol{\zeta}) &\triangleq  \argmax_{j \in [m-1]} \left( \left(\zeta_{(j+1) \bmod m}-\zeta_{j}\right) \bmod q \right) \notag
\end{align}
for a given cutting vector $\boldsymbol{\zeta}$.

\begin{example}
Let $n=4$, $q=7$, and $\bm{j}^{\rm gr} = (0,1,2,5)$. Then, the range of $\bm{j}^{\rm gr}$ is
\begin{displaymath}
\begin{split}
&\min\left( 0 - (1 - 7 \cdot 1),1-(2-7 \cdot 1),2-(5-7 \cdot 1), \right.\\
&\;\;\;\;\;\;\;\,  \left. 5-(0- 7 \cdot 0) \right) + 1 
= \min(6,6,4,5)+1 = 5.
\end{split}
\end{displaymath}
\end{example}

\subsection{The Case $m=3$}

\begin{lemma} \label{th:m3}
The minimum distance $d_{\rm opt}(q,3,L)$, for $L \geq 2$ and $q \geq 13$, is upper-bounded by $6$.
\end{lemma}

\begin{IEEEproof}
The proof is based on the template codeword from \cite[Theorem 4]{yan03}. For convenience of the reader, we restate the corresponding \emph{template support matrix} below
\begin{equation} \label{eq:templatem3}
\left[ \begin{smallmatrix}
0 & 0 & 2i-2k & 2i-2k & -2i & -2i \\
0 & -2i+k & 0 & -i & -i & -2i+k \\
0 & -4i+2k & -2i+2k & -4i+2k & 0 & -2i+2k \end{smallmatrix} \right]
\end{equation}
where $q \geq 5$, %
$i \in [q] \setminus \{0\}$, and $k \in [q]$ with $k \neq i,2i$, 
and where all operations are taken modulo $q$. The template support matrix is obtained by extracting the columns of the parity-check matrix corresponding to the support set of the underlying codeword. When constructing this matrix, the integer representation of the columns from (\ref{eq:form}) is used.

Now, a column of the general form $(x,x+y,x+2y)^T \bmod q$, $x,y \in [q]$, in this matrix has column group index of $y$. Thus, the sequence of column group indices corresponding to the template matrix in (\ref{eq:templatem3}) is
\begin{displaymath}
\left(0,-2i+k,-2i+2k,-3i+2k,i,k \right) \bmod q.
\end{displaymath}
For $i=1$ and $k=0$, we get the following sorted sequence
\begin{equation} \label{eq:groupindexseqm3}
\left(-3,-2,-2,0,0,1 \right) \bmod q.
\end{equation}

Since array LDPC codes are quasi-cyclic with period $q$, \emph{cyclically incrementing} a sequence of column group indices $\bm{j}^{\rm gr}=(j_0^{\rm gr},\dots,j_{n-1}^{\rm gr})$ of length $n$ of a codeword in an array LDPC code by an integer $\kappa \geq 1$ results in the valid sequence $(j_0^{\rm gr}+\kappa,\dots,j_{n-1}^{\rm gr}+\kappa) \bmod q$ of column group indices of a codeword that is obtained by cyclically shifting the given codeword by $\kappa q$ positions to the right.

If the range of a sorted (in nondecreasing order) sequence of column group indices corresponding to a codeword in the uncoupled array LDPC code is at most $\epsilon(\boldsymbol{\zeta})$, it can always be cyclically incremented so that the corresponding codeword has coordinates with column group index $j$ within $\left[\zeta_{j_{\rm max}(\boldsymbol{\zeta})}, \zeta_{j_{\rm max}(\boldsymbol{\zeta})+1} \right)$ when $j_{\rm max}(\boldsymbol{\zeta}) < m-1 = 2$, and within $\left[\zeta_{m-1}, q \right) \cup \left[0, \zeta_{0} \right)$ when $j_{\rm max}(\boldsymbol{\zeta}) = m-1 = 2$. 
 Consequently, there will exist a codeword in the coupled code (for all $L \geq 2$) with all coordinates within the same column section when $j_{\rm max}(\boldsymbol{\zeta}) < m-1 = 2$, or within two consecutive column sections when $j_{\rm max}(\boldsymbol{\zeta}) = m-1 = 2$. Thus, since $\min_{\boldsymbol{\zeta}} \epsilon(\boldsymbol{\zeta}) =  \left \lceil \frac{q}{m} \right \rceil = \left \lceil \frac{q}{3} \right \rceil$, there will exist a codeword in the SC-LDPC code of weight $6$ for all cutting vectors $\boldsymbol{\zeta}$ as long as
\begin{equation} \notag
 \left \lceil \frac{q}{m} \right \rceil = \left \lceil \frac{q}{3} \right \rceil \geq \spn\left(\left(-3,-2,-2,0,0,1\right) \bmod q\right).
\end{equation}
Since $\spn((-3,-2,-2,0,0,1) \bmod q) = 5$ for $q \geq 7$ and $4$ for $q=5$, 
the smallest $q$ that satisfies this inequality is $q=13$, and the result follows since the range of (\ref{eq:groupindexseqm3}) for $q \geq 13$ is constant.
\end{IEEEproof}

\begin{example}
Consider the case of $q=7$, $m=3$, and the template support matrix in (\ref{eq:templatem3}) for $i=1$ and $k=0$. The sorted (in nondecreasing order) sequence of column group indices is
$(0,0,1,4,5,5)$ which has range $5$. By cyclically incrementing this sequence by two, we get $(0,0,2,2,3,6)$. The corresponding support matrices (both of which correspond to codewords) are
\begin{displaymath}
\left[ \begin{smallmatrix}
0 & 0 & 2 & 2 & 5 & 5 \\
0 & 5 & 0 & 6 & 6 & 5 \\
0 & 3 & 5 & 3 & 0 & 5 \end{smallmatrix} \right] \text{ and }
\left[ \begin{smallmatrix}
0 & 0 & 2 & 2 & 5 & 5 \\
2 & 0 & 2 & 1 & 1 & 0 \\
4 & 0 & 2 & 0 & 4 & 2 \end{smallmatrix} \right]
\end{displaymath}
respectively. For the cutting vector $\boldsymbol{\zeta}=(4,5,6)$, we have $j_{\rm max}(\boldsymbol{\zeta})=m-1=2$, $\epsilon(\boldsymbol{\zeta})=5$, $\left[\zeta_{m-1}, q \right) \cup \left[0, \zeta_{0} \right) = \left[6,7\right) \cup \left[0,4\right) = \{0,1,2,3,6\}$,  and
\begin{displaymath}
\left[ \begin{smallmatrix}
\bm{H}_0 &  \\
\bm{H}_1 & \bm{H}_0 \\
& \bm{H}_1 
\end{smallmatrix} \right]
= \left[ \begin{mysmallmatrix}
\bm{I} & \bm{I} & \bm{I} & \bm{I} & \bm{0} & \bm{0} & \cellcolor{LightCyan}{\bm{0}} & \vline & & & & & & & \\
\bm{I} & \bm{P} & \bm{P}^2 & \bm{P}^3 & \bm{P}^4 & \bm{0} & \bm{0} & \vline & & & & & & & \\
\bm{I} & \bm{P}^2 & \bm{P}^4 & \bm{P}^6 & \bm{P} & \bm{P}^3 & \bm{0} & \vline & & & & & & & \\
\\[-0.5ex]\hline \\[+0.5ex]
\bm{0} & \bm{0} & \bm{0} & \bm{0} & \bm{I} & \bm{I} & \color{blue}{\bm{I}} & \vline & \color{blue}{\bm{I}} & {\bm{I}} & \color{blue}{\bm{I}} & \color{blue}{\bm{I}} & \bm{0} & \bm{0} & \bm{0} \\
\bm{0} & \bm{0} & \bm{0} & \bm{0} & \bm{0} & \bm{P}^5 & \color{blue}{\bm{P}^6} & \vline & \color{blue}{\bm{I}} & {\bm{P}} & \color{blue}{\bm{P}^2} & \color{blue}{\bm{P}^3} & \bm{P}^4 & \bm{0} & \bm{0} \\
\bm{0} & \bm{0} & \bm{0} & \bm{0} & \bm{0} & \bm{0} & \color{blue}{\bm{P}^5} & \vline & \color{blue}{\bm{I}} & {\bm{P}^2} & \color{blue}{\bm{P}^4} & \color{blue}{\bm{P}^6} & \bm{P} & \bm{P}^3 & \bm{0} \\
\\[-0.5ex]\hline \\[+0.5ex]
& & & & & & &  \vline & \bm{0} & \bm{0} & \bm{0} & \bm{0} & \bm{I} & \bm{I} & \bm{I} \\
& & & & & & & \vline & \bm{0} & \bm{0} & \bm{0} & \bm{0} & \bm{0} & \bm{P}^5 & \bm{P}^6 \\
& & & & & & & \vline & \bm{0} & \bm{0} & \bm{0} & \bm{0} & \bm{0} & \bm{0} & \bm{P}^5
\end{mysmallmatrix}  \right].
\end{displaymath}
From the highlighted blue columns in the matrix above we can identify a weight-$6$ codeword corresponding to the sequence $(0,0,2,2,3,6)$  of column group indices mentioned above. For other cutting vectors like $\zeta=(3,4,6)$, we get $\epsilon(\boldsymbol{\zeta})=4$, which is less than the range, and there will be no codewords of the type in (\ref{eq:templatem3}) with $i=1$ and $k=0$ (there are for $i=2$ and $k=0$) in the coupled code.
\end{example}

By combining Lemma~\ref{th:m3} with \cite[Corollary 6]{mit14}, we get the following theorem.

\begin{theorem} \label{th:m3equal}
For $L \geq 2$ and $q \geq 13$, the minimum distance $d_{\rm opt}(q,3,L) = 6$.
\end{theorem}

\subsection{The Case $m=4$}

\begin{lemma} \label{th:m4}
The minimum distance $d_{\rm opt}(q,4,L)$, for $L \geq 2$ and $q \geq 41$, is upper-bounded by $10$.
\end{lemma}

\begin{IEEEproof}
The proof follows the same main idea of the proof of Lemma~\ref{th:m3} above using the template support matrix ($q \geq 11$)
\begin{equation} \notag %
\left[ \begin{smallmatrix}
0 & 0 & -16 & -16 & -13 & -13 & -9 & -9 & -1 & -1 \\
0 & -2 & -10 & -8 & -10 & -6 & -8 & -6 & 0 & -2 \\
0 & -4 & -4 & 0 & -7 & 1 & -7 & -3 & 1 & -3 \\
0 & -6 & 2 & 8 & -4 & 8 & -6 & 0 & 2 & -4 \end{smallmatrix} \right]
\end{equation}
which we have found using the algorithm from \cite{ros14}. The remaining technical details are omitted for brevity.
\end{IEEEproof}

Note that the template support matrix for $m=4$ given in Fig.~3 in \cite{sug08} will fail to prove the above result, since the sorted  (in nondecreasing order) sequence of column group indices has range $19$ for $q \geq 23$. As a consequence, the template matrix in Fig.~3 in \cite{sug08} can only prove an upper bound of $10$ for $q \geq 73$.

\begin{proposition}
There is a codeword of weight $10$ in $\mathcal{C}(q,4,L,\boldsymbol{\zeta})$, $q \geq 13$, for all positive integers $L \geq 2$ and cutting vectors $\boldsymbol{\zeta}$ if one of the following conditions are true.
\begin{enumerate} \label{prop:1}
\item $\zeta_1-\zeta_0 \geq 2$ and $\zeta_2-\zeta_1 \geq 9$.
\item $\zeta_1-\zeta_0 \geq 9$ and $\zeta_2-\zeta_1 \geq 2$.
\item $\zeta_3-\zeta_2 \geq 2$ and $q+\zeta_0-\zeta_3 \geq 9$.
\item $\zeta_3-\zeta_2 \geq 9$ and $q+\zeta_0-\zeta_3 \geq 2$.
\item $\max(\zeta_{1}-\zeta_0, \zeta_2-\zeta_1,\zeta_3-\zeta_2,q+\zeta_0-\zeta_3) \geq 11$.
\end{enumerate}
\end{proposition}

\begin{IEEEproof}
The proof is omitted due to lack of space.
\end{IEEEproof}

\begin{lemma} \label{lem:1}
The minimum distances $d_{\rm opt}(31,4,L)$ and $d_{\rm opt}(37,4,L)$, for $L \geq 2$, are both upper-bounded by $10$.
\end{lemma}

\begin{IEEEproof}
For the case $q=31$, the number of cutting vectors that satisfy none of the five conditions in Proposition~\ref{prop:1} is only $35$, while the total number of possible cutting vectors is $35\,960$. Thus, by running the algorithm from \cite{ros12}, adapted to the case of SC codes, on these $35$ cases we have verified that there are indeed codewords of weight $10$ for all cutting vectors, and the result follows.

For $q=37$, there are no cutting vectors that satisfy none of the five conditions in Proposition~\ref{prop:1}, from which it follows that $d_{\rm opt}(37,4,L) \leq 10$ for all positive integers $L \geq 2$.
\end{IEEEproof}

By combining Lemmas~\ref{th:m4} and \ref{lem:1}, \cite[Theorem 5]{mit14}, and \cite[Corollary~4.2]{sug08}, we get the following theorem.

\begin{theorem} \label{th:m4real}
For $L \geq 2$ and $q \geq 31$, the minimum distance $d_{\rm opt}(q,4,L)=10$.
\end{theorem}

\subsection{The Case $m=5$}

\begin{theorem} \label{th:m5}
The minimum distance $d_{\rm opt}(q,5,L)$, for $L \geq 2$ and $q \geq 59$, is upper-bounded by $12$.
\end{theorem}

\begin{IEEEproof}
The proof follows the same main idea of the proofs of Lemmas~\ref{th:m3} and \ref{th:m4} above using the template support matrix from Fig.~4 in \cite{sug08}, and is omitted for brevity.
\end{IEEEproof}

Note that there appears to be no equivalent to Proposition~\ref{prop:1} (except for the last condition) for $m=5$, since the template support matrix from Fig.~4 in \cite{sug08} (which is used for the proof of Theorem~\ref{th:m5}) does not have the required structure. Also, the algorithm from \cite{ros14} was not able to identify other nonequivalent template support matrices with a suitable structure. This has also been ``confirmed'' by the fact that for $q=53$ we have been able to identify a cutting vector that gives a minimum distance of $14$ (see Table~\ref{table:SCarrayLDPC}).

\begin{theorem} \label{th:m5_1}
The minimum distance $d_{\rm opt}(q,5,L)$, for $L \geq 2$ and $q \geq 29$, is upper-bounded by $16$.
\end{theorem}

\begin{IEEEproof}
The proof follows the same main idea of the proofs of Lemmas~\ref{th:m3} and \ref{th:m4} and Theorem~\ref{th:m5} above using a template support matrix which was found using the algorithm from \cite{ros14}. Due to lack of space, the actual template support matrix and the technical details of the proof are omitted.
\end{IEEEproof}

\section{Numerical Results}

In this section, we present some numerical results for the cases $m=3$, $4$, and $5$.

\subsection{The Case $m=3$}

By performing an exhaustive search over all cutting vectors $\boldsymbol{\zeta}$ the optimal minimum distance $d_{\rm opt}(q,3)$ was determined for $5 \leq q \leq 11$ (see Table~\ref{table:SCarrayLDPC}). For each of these values of $q$, the $d_{\rm min}$-optimal cutting vector is not unique, i.e., there are several cutting vectors that give the best minimum distance. For instance, for $q=5$, we found (for $L=10$) the $h_{\rm min}$-optimal (where $h_{\rm min}$ denotes stopping distance) cutting vectors displayed in Table~\ref{table:q5m3cutting}. For each cutting vector, in the second and fourth columns we have tabulated the corresponding minimum and stopping distance, respectively. The corresponding multiplicities are tabulated in the third (minimum distance) and fifth (stopping distance) columns. Note that the \emph{optimal} cutting vector $\boldsymbol{\zeta}=(2,3,5)$ for $(3,3)$ and $(4,2)$ absorbing sets from \cite{mit14} is also optimal when it comes to minimum/stopping distance, since it gives the optimal  minimum/stopping distance of $10$. However, it gives slightly more stopping sets of size $10$ (see Table~\ref{table:q5m3cutting}). On the other hand, for $q=7$, the \emph{optimal} cutting vector $\boldsymbol{\zeta}=(2,4,6)$ for $(3,3)$ and $(4,2)$ absorbing sets from \cite{mit14} is \emph{not} optimal when it comes to minimum distance, since it gives a minimum distance of only $6$ (the optimal minimum distance is $8$).  For $q=5,7$, $d_{\rm min}$-optimal cutting vectors that also give the lowest possible minimum distance multiplicity, denoted as $(d_{\rm min},N_{d_{\rm min}})$-optimal cutting vectors, are displayed within the parentheses in the sixth column of Table~\ref{table:SCarrayLDPC} (first and second row, respectively).

For $q \geq 13$, it follows from Theorem~\ref{th:m3equal} that the optimal minimum distance is $6$. %
Also, all cutting vectors give a minimum distance of $6$, but possibly with different multiplicities.

\begin{table}[tbp]
\scriptsize \centering \caption{Minimum/Stopping Distance Results for Array-Based SC-LDPC Codes for $q=5$, $m=3$, $L=10$, and Different $h_{\rm min}$-Optimal Cutting Vectors}\label{table:q5m3cutting}
\def\Hline{\noalign{\hrule height 2\arrayrulewidth}}
\vskip -3.0ex %
\begin{tabular}{ccccc}
\Hline \\ [-2.0ex]
   $\boldsymbol{\zeta}$ & $d(5,3,10,\boldsymbol{\zeta})$ & Mult. & $h(5,3,10,\boldsymbol{\zeta})$ & Mult. \\
\hline
\\ [-2.0ex] \hline  \\ [-2.0ex]
$(0,1,3)$ &  10 & 20 & 10 & 65 \\
$(0,2,3)$ &  10 & 20 & 10 & 65 \\
$(1,2,4)$ &  10 & 19 & 10 & 59 \\
$(1,3,4)$ &  10 & 19 & 10 & 59 \\
$(2,3,5)$ &  10 & 20 & 10 & 65 \\
$(2,4,5)$ &  10 & 20 & 10 & 65 \end{tabular}
\end{table}

\subsection{The Case $m=4$}

By performing an exhaustive search over all cutting vectors $\boldsymbol{\zeta}$  (with the help of Proposition~\ref{prop:1} to reduce the search space when $q > 11$) the optimal minimum distance $d_{\rm opt}(q,4)$ was determined for $5 \leq q \leq 29$. For each of these values of $q$, the $d_{\rm min}$-optimal cutting vector is not unique, i.e., there are several cutting vectors that give the best minimum distance. For instance, for $q=7$, we found (for $L=10$) the $h_{\rm min}$-optimal cutting vectors displayed in Table~\ref{table:q7m4cutting}. For each cutting vector, in the second and fourth columns we have tabulated the corresponding minimum and stopping distance, respectively. The corresponding multiplicities are tabulated in the third (minimum distance) and fifth (stopping distance) columns. For $5 \leq q \leq 29$, the optimal minimum distance and corresponding $(d_{\rm min},N_{d_{\rm min}})$-optimal cutting vectors  (displayed within the parentheses) are given in the fourth column of Table~\ref{table:SCarrayLDPC} (the first eight rows, respectively). For $q \geq 31$, it follows from Theorem~\ref{th:m4real} that $d_{\rm opt}(q,4)=10$.

\begin{table}[tbp]
\scriptsize \centering \caption{Minimum/Stopping Distance Results for Array-Based SC-LDPC Codes for $q=7$, $m=4$, $L=10$, and Different $h_{\rm min}$-Optimal Cutting Vectors}\label{table:q7m4cutting}
\def\Hline{\noalign{\hrule height 2\arrayrulewidth}}
\vskip -3.0ex %
\begin{tabular}{ccccc}
\Hline \\ [-2.0ex]
   $\boldsymbol{\zeta}$ & $d(7,4,10,\boldsymbol{\zeta})$ & Mult. & $h(7,4,10,\boldsymbol{\zeta})$ & Mult. \\
\hline
\\ [-2.0ex] \hline  \\ [-2.0ex]
$(0,2,3,5)$ &  14 & 30 & 14 & 401 \\
$(0,2,4,6)$ &  14 & 30 & 14 & 695 \\
$(1,3,4,6)$ &  14 & 29 & 14 & 393 \\
$(1,3,5,7)$ &  14 & 30 & 14 & 695 \\
$(2,4,5,7)$ &  14 & 30 & 14 & 401 \end{tabular}
\end{table}

\begin{table*}[t]
\scriptsize \centering \caption{Optimal Minimum Distance Results for Array-Based SC-LDPC Codes for Different Values of $q$ and $m$. Values in Bold Are New Results, While Non-Bold Values Are Taken From the Literature. The Results Are Compared With Those of Array-Based Uncoupled LDPC Codes Taken From \cite{yan03,sug08,ros14}. Within the Parentheses the Corresponding Cutting Vectors Are Displayed. Cutting Vectors in Italics Are $d_{\rm min}$-Optimal, While  Cutting Vectors in Bold Are $(d_{\rm min},N_{d_{\rm min}})$-Optimal}\label{table:SCarrayLDPC}
\def\Hline{\noalign{\hrule height 2\arrayrulewidth}}
\vskip -3.0ex %
\begin{tabular}{clclclc}
\Hline \\ [-2.0ex]
   $q$ & $d_{\rm opt}(q,5)$ & $d(q,5)$ & $d_{\rm opt}(q,4)$ & $d(q,4)$ & $d_{\rm opt}(q,3)$ & $d(q,3)$ \\
\hline
\\ [-2.0ex] \hline  \\ [-2.0ex]
5 & &  & $\bm{10}$ $\bf (1,2,3,4)$ & $8$ & $\bm{10}$ $\bf (1,2,4)$ & $6$ \\
7 & $\bm{14}$ $\bf (1,2,3,4,6)$ & 12 & $\bm{14}$ $\bf (1,3,4,6)$ & $8$ & $\bm{8}$ $\bf (1,2,5)$ & $6$ \\
11 & $\bm{22}$ $\it (0,1,2,5,8)$ & 10 &  $\bm{14}$ $\bf (1,3,6,8)$ & $10$ & $\bm{6}$ & $6$ \\
13 & $\bm{24}$ $\it (0,2,5,7,10)$ & 12 &  $\bm{12}$ $\bf (0,3,7,9)$ & $10$ & $\bm{6}$ & $6$ \\
17 & $\bm{20}$, $\bm{22}$, or $\bm{24}$ $(0,4,7,11,14)$ & 12 & $\bm{12}$ $\bf (2,5,10,13)$ & $10$ & $\bm{6}$ & $6$ \\
19 & $\bm{20}$ or $\bm{22}$ %
$(0,4,8,12,16)$ & 12 & $\bm{12}$ $\bf (3,7,12,16)$ & $10$ & $\bm{6}$ & $6$ \\
23 & $\bm{20}$ %
$\it (5,10,15,19,23)$ & 12 & $\bm{12}$ $\bf (4,8,15,19)$ & $10$ & $\bm{6}$ & $6$ \\
29 & $\bm{16}$ $\it (8,11,18,26,29)$  & 12 & $\bm{12}$ $\bf (4,11,18,25)$ & $10$ & $\bm{6}$ & $6$ \\
31 & $\bm{16}$ $\it (9,15,21,24,31)$ & 12 & $\bm{10}$  & $10$ & $\bm{6}$ & $6$ \\
37 & $\bm{16}$ $\it (0,10,19,28,30)$ & 12 & $\bm{10}$  & $10$ & $\bm{6}$ & $6$ \\
41 & $\bm{16}$ $\it (0,9,18,25,32)$ & 12 & $\bm{10}$  & $10$ & $\bm{6}$ & $6$ \\
43 & $\bm{16}$ $\it (0,8,19,24,33)$ & 12 & $\bm{10}$  & $10$ & $\bm{6}$ & $6$ \\
47 & $\bm{16}$ $\it (7,16,26,36,47)$ %
    & 12 & $\bm{10}$  & $10$ & $\bm{6}$ & $6$ \\
53 & $\bm{14}$ $\bf (0,11,22,33,42)$ & 12 & $\bm{10}$  & $10$ & $\bm{6}$ & $6$ \\
$59 - 79$ & $\bm{12}$ & 12 & $\bm{10}$  & $10$ & $\bm{6}$ & $6$ \\
$> 79$ & $\bm{10}$ or $\bm{12}$ & 10 or 12 & $\bm{10}$  & $10$ & $\bm{6}$ & $6$ 
\end{tabular}
\vspace{-1mm}
\end{table*}

\subsection{The Case $m=5$}

For $q=7$, we have performed an exhaustive search over all possible cutting vectors. The optimal minimum distance $d_{\rm opt}(7,5)$ is $14$ and a $(d_{\rm min},N_{d_{\rm min}})$-optimal cutting vector is displayed within the parentheses in the second column of  Table~\ref{table:SCarrayLDPC} (second row). For $q=11$ and $13$, the optimal minimum distance is as high as $22$ and $24$, respectively, and $d_{\rm min}$-optimal cutting vectors  are shown within the parentheses in the second column of  Table~\ref{table:SCarrayLDPC} (third and fourth row, respectively). For $q=17$ and $19$, upper and lower bounds on the optimal minimum distance, tabulated in the second column of Table~\ref{table:SCarrayLDPC}, have been established from an exhaustive search over all cutting vectors. The displayed values are bounds since we were not able to exhaustively enumerate all codewords of weight at most $\tau$, when $\tau \geq 20$, for a given cutting vector, for these two values of $q$.  Within the parentheses a corresponding cutting vector (which establishes the upper bound) is also tabulated. The lower bounds were determined by running the algorithm from \cite{ros12}, adapted to the case of SC codes, for the specific cutting vectors displayed within the parentheses. %
For $23 \leq q \leq 53$, the tabulated values (in the second column of  Table~\ref{table:SCarrayLDPC}) are again the exact values of $d_{\rm opt}(q,5)$ and also  (within the parentheses) $d_{\rm min}$-optimal ($(d_{\rm min},N_{d_{\rm min}})$-optimal for $q=53$) cutting vectors are displayed. For $59 \leq q \leq 79$,  $d_{\rm opt}(q,5) = 12$, which follows from Theorem~\ref{th:m5}, \cite[Table~I]{ros14}, and Theorem~5 in \cite{mit14}. For $q > 79$, it follows from Theorem~\ref{th:m5}, Corollary~4.4 in \cite{sug08}, and Theorem~5 in \cite{mit14} that $d_{\rm opt}(q,5)$ is either $10$ or $12$, although we conjecture it to be $12$.

\section{Conclusion and Future Work}

In this work, we have studied in detail the minimum distance of array-based SC-LDPC codes. Several tight upper bounds on the optimal minimum distance for coupling length $L \geq 2$ and $m=3,4,5$, that are independent of $q$ and that are valid for all values of $q \geq  q_0$  where $q_0$ depends on $m$, have been presented. Furthermore, we have conducted an exhaustive search %
over all cutting vectors for small values of $q$ ($m=3,4,5$) which shows that by carefully selecting the cutting vector, the minimum distance (when $q$ is not very large) can be significantly increased, especially for $m=5$.

An interesting topic for future work is to consider the correlation with absorbing sets. In particular, will a $(d_{\rm min},N_{d_{\rm min}})$-optimal cutting vector also be \emph{close-to-optimal} when it comes to problematic absorbing sets, and/or vice-versa?

\balance


\begin{thebibliography}{10}
\providecommand{\url}[1]{#1}
\csname url@samestyle\endcsname
\providecommand{\newblock}{\relax}
\providecommand{\bibinfo}[2]{#2}
\providecommand{\BIBentrySTDinterwordspacing}{\spaceskip=0pt\relax}
\providecommand{\BIBentryALTinterwordstretchfactor}{4}
\providecommand{\BIBentryALTinterwordspacing}{\spaceskip=\fontdimen2\font plus
\BIBentryALTinterwordstretchfactor\fontdimen3\font minus
  \fontdimen4\font\relax}
\providecommand{\BIBforeignlanguage}[2]{{%
\expandafter\ifx\csname l@#1\endcsname\relax
\typeout{** WARNING: IEEEtran.bst: No hyphenation pattern has been}%
\typeout{** loaded for the language `#1'. Using the pattern for}%
\typeout{** the default language instead.}%
\else
\language=\csname l@#1\endcsname
\fi
#2}}
\providecommand{\BIBdecl}{\relax}
\BIBdecl

\bibitem{bal14}
M.~Baldi, G.~Cancellieri, and F.~Chiaraluce, ``Array convolutional low-density
  parity-check codes,'' \emph{IEEE Commun.~Lett.}, vol.~18, no.~2, pp.
  336--339, Feb. 2014.

\bibitem{mit14}
D.~G.~M. Mitchell, L.~Dolecek, and {D. J. Costello, Jr.}, ``Absorbing set
  characterization of array-based spatially coupled {LDPC} codes,'' in
  \emph{Proc.~IEEE Int.~Symp.~Inf.~Theory (ISIT)}, Honolulu, HI, Jun./Jul.
  2014, pp. 886--890.

\bibitem{fan00}
J.~L. Fan, ``Array codes as low-density parity-check codes,'' in \emph{Proc.
  2nd Int. Symp. Turbo Codes {\&} Rel. Topics}, Brest, France, Sep. 2000, pp.
  543--546.

\bibitem{fel99}
A.~J. Felstr\"{o}m and K.~S. Zigangirov, ``Time-varying periodic convolutional
  codes with low-density parity-check matrix,'' \emph{IEEE Trans.~Inf.~Theory},
  vol.~45, no.~6, pp. 2181--2191, Sep. 1999.

\bibitem{kud13}
S.~Kudekar, T.~Richardson, and R.~L. Urbanke, ``Spatially coupled ensembles
  universally achieve capacity under belief propagation,'' \emph{IEEE
  Trans.~Inf.~Theory}, vol.~59, no.~12, pp. 7761--7813, Dec. 2013.

\bibitem{mit02}
T.~Mittelholzer, ``Efficient encoding and minimum distance bounds of
  {R}eed-{S}olomon-type array codes,'' in \emph{Proc.~IEEE
  Int.~Symp.~Inf.~Theory (ISIT)}, Lausanne, Switzerland, Jun./Jul. 2002, p.
  282.

\bibitem{yan03}
K.~Yang and T.~Helleseth, ``On the minimum distance of array codes as {LDPC}
  codes,'' \emph{IEEE Trans.~Inf.~Theory}, vol.~49, no.~12, pp. 3268--3271,
  Dec. 2003.

\bibitem{sug08}
K.~Sugiyama and Y.~Kaji, ``On the minimum weight of simple full-length array
  {LDPC} codes,'' \emph{IEICE Trans.\ Fundamentals}, vol. E91-A, no.~6, pp.
  1502--1508, Jun. 2008.

\bibitem{esm09}
M.~Esmaeili and M.~J. Amoshahy, ``On the stopping distance of array code
  parity-check matrices,'' \emph{IEEE Trans.~Inf.~Theory}, vol.~55, no.~8, pp.
  3488--3493, Aug. 2009.

\bibitem{ros14}
E.~Rosnes, M.~A. Ambroze, and M.~Tomlinson, ``On the minimum/stopping distance
  of array low-density parity-check codes,'' \emph{IEEE Trans.~Inf.~Theory},
  vol.~60, no.~9, pp. 5204--5214, Sep. 2014.

\bibitem{dol10}
L.~Dolecek, Z.~Zhang, V.~Anantharam, M.~J. Wainwright, and B.~Nikolic,
  ``Analysis of absorbing sets and fully absorbing sets of array-based {LDPC}
  codes,'' \emph{IEEE Trans.~Inf.~Theory}, vol.~56, no.~1, pp. 181--201, Jan.
  2010.

\bibitem{olc03}
S.~\"Ol\c{c}er, ``Decoder architecture for array-code-based {LDPC} codes,'' in
  \emph{Proc.\ IEEE Global Telecommun. Conf. (GLOBECOM)}, vol.~4, San
  Francisco, CA, Dec. 2003, pp. 2046--2050.

\bibitem{bha05}
P.~Bhagawat, M.~Uppal, and G.~Choi, ``{FPGA} based implementation of decoder
  for array low-density parity-check codes,'' in \emph{Proc.\ IEEE Int. Conf.
  Acoustics, Speech, and Signal Processing (ICASSP)}, vol.~5, Philadelphia, PA,
  Mar. 2005, pp. 29--32.

\bibitem{pus11}
A.~E. Pusane, R.~Smarandache, P.~O. Vontobel, and {D. J. Costello, Jr.},
  ``Deriving good {LDPC} convolutional codes from {LDPC} block codes,''
  \emph{IEEE Trans.~Inf.~Theory}, vol.~57, no.~2, pp. 835--857, Feb. 2011.

\bibitem{ros12}
E.~Rosnes, {\O}.~Ytrehus, M.~A. Ambroze, and M.~Tomlinson, ``Addendum to ``{An}
  efficient algorithm to find all small-size stopping sets of low-density
  parity-check matrices'','' \emph{IEEE Trans.~Inf.~Theory}, vol.~58, no.~1, pp.
  164--171, Jan. 2012.

\end{thebibliography}
\end{document}